\documentclass[aps,twocolumn,secnumarabic,showpacs,nobalancelastpage,amsmath,amssymb,
nofootinbib]{revtex4}
\usepackage{graphics}      
\usepackage{graphicx}      
\usepackage{longtable}     
\usepackage{url}           
\usepackage{bm}

\begin{document}
\title{An Induced Representation Method for Studying the Stability of Saturn's Ring}
\author         {Soumangsu Bhusan Chakraborty}
\email          {soumangsu.bhusan-chakraborty@polytechnique.edu}

\affiliation    {Ecole Polytechnique, Palaiseau 91120, France}
\author         {Siddhartha Sen}
\email          {siddhartha.sen@tcd.ie} 
\affiliation    {CRANN, Trinity College Dublin, Dublin  2, Ireland}
\date{\today}

\begin{abstract}
Using the method of induced representation for groups MacKey proved 
 a set of formulas that could be used to calculate the sum of  powers of the 
 eigenvalues of a matrix which  was symmetric under a
 finite group. We use MacKey's results
 to derive the stability condition, $m>Cn^3$,  for
a ring of Saturn model due to Maxwell where $n$ is
the number of unit mass particles in a ring and $m$
the mass of Saturn. In Maxwell's model a ring of Saturn
is considered to be a symmetrically arranged  collection of
$n$ identical unit mass particles revolving round Saturn
in a circular orbit.
\end{abstract}

\pacs{45.20J,45.10Hj,0220-a}

\maketitle
\section{Introduction}
In this paper a group theoretical result of  MacKey   \cite{Sternberg}
that relates the sum of  powers of the eigenvalues of a
perturbation matrix, symmetric under a finite group, to group characters is
 used to derive Maxwell's stability condition $m>C n^3$ for a ring of Saturn.
 \cite{maxwell}  where $C$ is a constant, $m$ is the mass of Saturn and 
 $n$ represents the number of identical
 symmetrically arranged unit mass particles that were assumed to 
 be revolving  round Saturn in a circular orbit by Maxwell.
 
 Using group theoretical methods for this stability problem
is reasonable as the $n$ identical particles
 that form a ring have  discrete rotational symmetry. The 
 effect of linear perturbation on such a system can be described in terms
 of the geometric picture of a vector bundles. In a vector bundle two spaces are joined
together in a precise way to form a larger space.For the Saturn's ring perturbation
problem these two spaces are: the initial symmetric location of the ring particles
on a circle and the space of perturbation attached to each particle.
The symmetry group of location induces a group action on the perturbation
space.  We will briefly outline the basic terminology of vector bundles
and induced representations that we need in order to
write down MacKey's formulas and use them to get the stability result stated. 
 
As stated already Maxwell assumed that each ring of Saturn
 was made up of $n$ identical unit mass particles all revolving around 
a central massive planet (Saturn) of mass $m$ in a  planar circular orbit
 with the same angular velocity. He had earlier in the work established that a liquid 
ring could not be stable and that a solid ring could not be stable if the
mass distribution was uniform. The stability condition  for the
 orbiting ring particles he proved was\cite{maxwell} 
$$m>0.43 n^3,$$. 

Recently  Maxwell's model has been revisited by many authors using
 different mathematical techniques \cite{roberts}\cite{moeckel1}\cite{princeton}\cite{pendes}\cite{solo}\cite{vinh}\cite{moeckel2}. 
 In these works, different ways of  analyzing the  linearized  $4n\times 4n$ dimensional perturbation matrix  appropriate for the 
 Saturn's ring problem are considered.  Simplification  of the problem follow once invariant subspace of the perturbation 
matrix are found which allow a projection of the perturbation matrix 
on to a particular invariant space. This is then shown to 
 reduce the stability problem, for $n$ even, to that of determining the 
eigenvalues of a $4\times 4$ matrix. However the  invariant  subspaces are found 
by ad hoc methods as there is no general approach available for finding them. 

The  virtue of our group theoretical approach is that the required stability 
conditions are found in terms of $4\times 4$ matrices, without determining 
invariant subspaces and the basic result that for linear stability $m=Cn^3$
where $C$ is a constant, is, as we will show easily obtained. The aim of this paper 
is to establish this result and not to precisely determine the constant $C$
  
 We start by describing the set of MacKey's character formulas for an induced representation.

\section{The character formulas}
Analyzing the linear stability of a system requires knowledge of
the eigenvalues of  a perturbation matrix,  which is a linear 
operator $T$ acting on a finite dimensional vector space $W$. For our 
problem the linear operator $T$ of interest has a discrete symmetry.
MacKey's character formulas relates the sum of powers of 
the eigenvalues of $T$ to group theoretical structures present
 in the problem. The proof of these formulas 
 are given in Sternberg \cite{Sternberg}. 

Induced representations are representations of a group
generated from a subgroup. We can also think of induced representations
as representations of a group on the sections of a vector bundle from a knowledge
of the representation of the group on the base space of the bundle.
Let us briefly explain these terms. We recall that 
 a vector bundle is a way of joining together two spaces where precise rules for
joining the two spaces have to be given. Such a joining
of two spaces is natural for linear perturbative stability problems.
For example in the Saturn's rings problem the two spaces to be joined are: the
original set of equilibrium position and momenta coordinates (the base space) and the 
space $R^2 \times R^2$ (the fiber) that describes the planar perturbation from their 
equilibrium values of the position and momenta coordinates.  Assigning positions 
and momenta  for each perturbed particle in phase space is called a section of
the bundle.To describe a section the 
location of a point in phase space and the corresponding perturbed position and 
momenta values have to be given. It represents a point of the  
vector bundle relevant for the Saturn's ring problem.  In this geometrical
 language a perturbation is a linear map 
between sections of a vector bundle. The symmetry features of such a 
 bundle are found from the symmetry of the original system by using the 
method of induced representations.  

Let us now introduce the precise vocabulary needed to properly describe the 
character formulas that we will use. Suppose $M$ is a finite set (the base space) with a  vector 
space $E_x$,
 the fiber, associated with each point $x\in M$. The vector bundle $E$ over $M$ 
is then  the disjoint union of all of these 
vector spaces $E_x$. So we write $E=\sqcup_{x\in M}E_x$.  There is also a 
natural projection map $\pi : E\longrightarrow M$ given by $\pi (v)=x$ if $v\in E_x$. 
Its inverse $E_x=\pi^{-1}(x)$ is the vector space associated with the point $x \in M$.  
A section of $E$ is the function $f:M\longrightarrow E$ which assigns a vector 
$f(x)\in E_x$ to each $x\in M$. 
So we can see that the map $f$ satisfies $\pi\circ f=\text{identity}.$   For a 
fixed 
$x\in M$, the space of all sections $f(x)$ is given by $\Gamma(E_x)$ and 
$\Gamma(E)=\bigoplus_{x\in M}\Gamma(E_x).$  Thus the projection map tells us 
where a particular fiber is located while a section tells us about points in the 
vector bundle. The space of sections is not a vector space but it
 can be made into a vector space by 
introducing a suitable way of adding sections and multiplying  them by scalars 
\cite{Sternberg}. 

  Let $W$ be a vector space given 
by $\Gamma (E)$ and $E$ the vector bundle over the finite  set $M$ on
which group $G$  acts transitively. Let $H$ be a subgroup of $G$ and
$(\rho,W)$ be a representation of the group $G$ with irreducible characters 
$\chi_1,\chi_2,\cdots,\chi_s$. Let us consider that 
$W=W_1\oplus W_2\oplus\cdots \oplus W_s$ be canonical decompositions of the 
space $W$ where each $W_i$ can be $m_i$ copies of $i^{th}$
irreducible representations of $G$ of dimension $d_i$. Let $P_i$ be the 
projection operator on the space $W_i$. Let us consider a linear 
map $T\in \mathrm{Hom}_G(W,W)$. Then 
we have;
$TP_i=P_iTP_i.$
We would like to calculate the eigenvalues of $TP_i$. If 
$\lambda_1,\lambda_2, \cdots,\lambda_{d_{i}}$ are its  eigenvalues  then we can 
write:
$\frac{1}{d_i}\mathrm{tr} (TP_i)=(\lambda_1+\lambda_2+ \cdots +\lambda_{d_{i}}),$ where 
$d_i=\chi_i(e)$ is the dimension of the $i^{th}$ irreducible representations.
Similarly we can write: $\frac{1}{d_i}\mathrm{tr} 
(T^2P_i)=(\lambda_{1}^2+\lambda_{2}^2+\cdots+\lambda_{d_{i}}^2).$ Hence knowing 
$\mathrm{tr}(T^kP_{i})$ where $k$ 
is an integer, for $1\leq k \leq d_i$,  we can in principle, determine all the 
eigenvalues. Let us consider a subgroup $H$ of $G$ such that $M=G/H$. So $E$ is 
the vector 
bundle over $M=G/H$ induced from a representation $(\sigma,V)$ of $H.$ It is shown in 
~\cite{Sternberg} that,
\begin{equation}\label{EQUN1}
 (Tf)(a)=\frac{1}{\#H}\sum_{b\in G}t(a,b)f(b), \ \forall a\in G.
\end{equation}
where $t(a,b)\in \mathrm{Hom}(V,V)$ is a linear operator (a matrix) that sends each 
element of the vector space associated with the coset $aH$ to that 
associated with coset $bH$. It can be proved that for a given $T\in 
\mathrm{Hom}(\Gamma(E),\Gamma(E))$ we can uniquely determine $t\in \mathrm{Hom}(V,V).$  
We are now in a position to write the complete form of the character formula 
as:~\cite{Sternberg} 
\begin{equation}\label{CF}
 \mathrm{tr}(T^{k}P_i)=\frac{1}{\#H}\sum_{a\in G}\overline{\chi_i(a)}\mathrm{tr}(t^{k}(a)),
\end{equation}
where we denote $t(e,a)$ by $t(a)$ (a matrix) with $e$ as the identity element 
of $G$ and hence of $H$.
This is the result we use to discuss the stability of Saturn's ring problem.

\section{Stability problem: assumptions}
Let us now formulate the dynamical problem of stability that we consider.
Our ring consists of $n$ identical point particles of 
unit mass revolving in a plane around Saturn in a circular orbit of constant radius 
 with constant angular velocity. We also assume that the $n$ identical particles are 
symmetrically arranged about the central mass and that they all lie on a plane and
 use  labels  from $0$ to $n-1$ in clockwise or anticlockwise sense to denote the 
$n$ particles of the ring and the label $n$ for Saturn. We set the mass of each particle 
$m_i=1, \ \forall i\in \{0,1,2,...,n-1\}$ and set  the mass of  Saturn to  be $m$. \\
\indent  We consider the ringed system in isolation and thus only include 
inter-particle gravitational interaction and gravitational interaction of each 
identical particle with  Saturn. 
 
\subsection{Relative equilibrium of the $1+n$ body planet ring system}
We now formulate the perturbation problem following Roberts~\cite{roberts} and Moeckel~ 
\cite{moeckel1}\cite{moeckel2} and then recast it in a form which highlights its symmetry.
Let $\textbf{q}_i\in \mathbb{R}^2$ be the generalized coordinates of the planet 
and let $\textbf{p}_i\in \mathbb{R}^2$ be their generalized 
momentum. Let $\textbf{q}=(\textbf{q}_0,\textbf{q}_1,\cdots,\textbf{q}_n)\in 
\mathbb{R}^{2(n+1)}$. The distance between the $i^{th}$ and the $j^{th}$ 
particle be $r_{ij}=\| \textbf{q}_i-\textbf{q}_j \|$. Using Newton's law of 
motion and the inverse square law of gravitation we get the following equation;\cite{roberts}
\begin{equation}
 m_i\ddot{\textbf{q}}_i=\sum_{i\neq 
j=0}^{n}\frac{m_im_j(\textbf{q}_i-\textbf{q}_j)}{r_{ij}^3}=\frac{\partial 
U}{\partial \textbf{q}_i}.
\end{equation}
Where $U(\textbf{q})$ is the Newtonian potential of the system given by
 $U(\textbf{q})=\sum_{i<j}\frac{m_im_j}{r_{ij}}.$
The generalized momentum can be written as 
$\textbf{p}_{i}=m_i\dot{\textbf{q}}_{i}$ and let 
$\textbf{p}=(\textbf{p}_0,\textbf{p}_1,...,\textbf{p}_n)\in 
\mathbb{R}^{2(n+1)}$.
Hence the equation of motions can be written as;\cite{roberts}
\begin{equation}
 \begin{array}{rcl}
  \dot{\textbf{q}}   &=&  M^{-1}\textbf{p}        = \frac{\partial H}{\partial 
\textbf{p}}   \\
  \dot{\textbf{p}}   &=&  \nabla U(\textbf{q})    = - \frac{\partial H}{\partial 
\textbf{q}},
 \end{array}
\end{equation}
where $M=\text{diag}\{m_0,m_0,m_1,m_1,\cdots,m_n,m_n\}$ is a diagonal, 
$2(n+1)\times 2(n+1)$ matrix and $H(\textbf{q},\textbf{p})$ is the Hamiltonian 
of the system.\\
 Let us consider the ring isomorphism $\mathbb{C}\longrightarrow 
\mathbb{M}_2(\mathbb{R})$ given by:
$(a+ib)\longmapsto \begin{bmatrix}
                     a  & b  \\
                     -b & a
                    \end{bmatrix} \  \text{with} \ (a,b)\in \mathbb{R}^2
.$
With the above isomorphism in mind we can make the following change of 
coordinates:\cite{roberts}
\begin{equation}\label{EOM}
 \begin{array}{rcl}
  \textbf{x}_i &=& e^{i\omega t}\textbf{q}_i,  \\
  \textbf{y}_i &=& e^{i\omega t}\textbf{p}_i.
 \end{array}
\end{equation}
Here $\frac{2\pi}{\omega}$ is the common period of rotation of the identical 
particles about the central planet and
$$e^{i\omega t}\longmapsto \begin{bmatrix}
                            \cos(\omega t)  & \sin(\omega t)  \\
                            -\sin(\omega t) & \cos(\omega t)
                           \end{bmatrix}.
$$
Hence in this new coordinates system the equation of motion 
becomes;\cite{roberts}
\begin{equation}
 \begin{array}{rcl}\label{HE}
  \dot{\textbf{x}} &=& A\textbf{x}+M^{-1}\textbf{y}  \\
  \dot{\textbf{y}} &=& \nabla U(\textbf{x})+A\textbf{y}, 
 \end{array}
\end{equation}
where $A= \begin{bmatrix}
            i\omega \mathbb{I} & 0 \\
            0                  & 0
           \end{bmatrix}$ with $\mathbb{I}$ as the $2\times 2$ identity 
matrix.\\
In this new set of coordinate system we can write the Hamiltonian of the system 
as:\cite{roberts}
\begin{equation}
 H(\textbf{x},\textbf{y})=\frac{1}{2}\textbf{y}^{T}M^{-1}\textbf{y} - 
U(\textbf{x}) - \textbf{x}^{T} A \textbf{y}. 
\end{equation}
For equilibrium we must have $(\dot{\textbf{x}},\dot{\textbf{y}})=(0,0)$ This 
gives 
\begin{equation}\label{EEqn}
 \nabla U(\textbf{x})+\omega^2M\textbf{x}=0.
\end{equation}
As all the revolving point masses are taken to have unit mass we
set  $m_i=1 \ \text{for} \ i\in \{0,1,2,...,n-1\}$ and set the mass of Saturn, $m_n=m$. We also scale the 
radius of the circular orbit 
to be equal to one. This gives 
$\textbf{x}_j=(\cos\theta_j,\sin\theta_j)$, where 
$\theta_j=\frac{2\pi j}{n}$ for $j\in \{0,1,2,\cdots,n-1\}$.
A little bit of trigonometry gives $r_{ij}=2\sin(\theta_{ij}/2) \ \text{and} \ 
r_{ij}^2=2(1-\cos(\theta_{ij}))$, where $\theta_{ij}=\frac{2\pi (j-i)}{n}$ for 
$i,j\in \{0,1,2,\cdots,n-1\}$. 
This will give \cite{roberts} $r_{0j}=2\sin(\theta_{j}/2)  \ \text{and} \ 
r_{0j}^2=2(1-\cos(\theta_{j})).$
Substituting these expressions in the equation ~\eqref{EEqn} we get 
$\omega=\omega(m)$ as~\cite{roberts}
$\omega^2=m+\frac{1}{2}\sigma_n$ where 
$\sigma_n=\sum_{k=1}^{n-1}\frac{1}{r_{0k}}=\frac{1}{2}\sum_{k=1}^{n-1}\csc\frac{
\pi k}{n}$. This formula
shows that as the central mass increases, the period of rotation of the 
identical bodies decreases.

\subsection{Linear stability matrix}
Linearizing the system of equations ~\eqref{HE} we get the  stability matrix;
$ T=\begin{bmatrix}
    A                      &  M^{-1}\\
    D\nabla U(\textbf{x})  &  A 
   \end{bmatrix}$
where  $S=D\nabla U(\textbf{x})$ denotes the derivative of the gradient of the 
potential and is a $2(n+1)\times 2(n+1)$ matrix.
The matrix $S$ can be written as;
$S=
\begin{bmatrix}
 S_{00} & \cdots & S_{0n}  \\
 \vdots & \ddots & \vdots \\
 S_{n0} & \cdots & S_{nn}
\end{bmatrix}.
$
Here $S_{ij}$ is a $2\times 2$ matrix given by 
$S_{ij}=\frac{m_im_j}{r_{ij}^3}[\mathbb{I}_{2\times 2}- 3x_{ij}x_{ij}^T]$ 
~\cite{roberts} if $i\neq j$
and $S_{jj}=-\sum_{i\neq j}S_{ij}$, where 
$x_{ij}=\frac{\textbf{x}_j-\textbf{x}_i}{r_{ij}}$. Note that $S$ is a block 
symmetric matrix
i.e. $S_{ij}=S_{ji}$.\\
Now substituting the value of $\textbf{x}_i=(\cos\theta_i,\sin\theta_i)$ into 
the expression of $S_{ij}$ we get
$$S_{0j}=\frac{1}{2r_{0j}^3} \begin{bmatrix}
                             -1+3\cos(\theta_j) & 3\sin(\theta_j) \\
                             3\sin(\theta_j)   & -1-3\cos(\theta_j)
                            \end{bmatrix}
 \text{for} \ j\neq 0,n,$$
and 
$S_{0n}=m \begin{bmatrix}
            -2 & 0 \\
            0  & 1
           \end{bmatrix}.
$
Hence we have: 
               $ S_{00} = -\sum_{j\neq 0}S_{j0}   
                       =-\sum_{j\neq 0}S_{0j}    
                       = -\sum_{j= 1}^{n-1}S_{0j}-S_{0n}  .$                    
More generally we can write the other elements of the matrix $S$ as:
$$S_{ij}= \frac{1}{2r_{ij}^3}\begin{bmatrix}
           -1+3\cos({\theta_j+\theta_i})  &  3\sin(\theta_j+\theta_i)  \\
           3\sin(\theta_j+\theta_i)       &  -1-3\cos({\theta_j+\theta_i})
          \end{bmatrix} \ 
$$
$\text{for} \ i,j\neq n \ \text{and} \ i\neq j,$
$$S_{jn}=m \begin{bmatrix}
            1-3\cos^2\theta_j           &  -3\cos\theta_j\sin\theta_j \\
            -3\cos\theta_j\sin\theta_j   &  1-\sin^2\theta_j
           \end{bmatrix} \ \text{for} \ j\neq n.
$$

\subsection{Group Theoretical formulation of the stability problem}
We are now ready to highlight symmetry features present in the problem.
Our system  consists of $n$ identical particles, symmetrically placed
on a circle, shifting the angular positions of the  point particles on the 
circle by an angle $2\pi k/n$ for all integer $k$ leaves the system unchanged. 
Thus our system possesses a $\mathbb{Z}_n$ symmetry.  We proceed to exploit this 
symmetry using the
geometrical picture of vector bundles and sections.\\ 
\indent Every particle is associated with a vector space 
$V=\mathbb{R}^2\bigoplus \mathbb{R}^2$ one for the two positions and the other 
for the two 
momentum. Now we know that $\mathbb{R}^2\bigoplus \mathbb{R}^2$ is isomorphic to 
$\mathbb{C}\bigoplus \mathbb{C}=\mathbb{C}^2.$ Let us denote the 
vector spaces associated with the $k^{th}$ particle as $E_k=\mathbb{C}\bigoplus 
\mathbb{C}$ and for each $\mathbb{C}$ we have a representation of 
the form $\exp(2\pi ik/n)$. So the representation $\rho_k$ over the space $E_k$ 
is given by the following matrix:
\begin{equation}
\rho_k = \begin{bmatrix}
 \exp(2\pi ik/n) & 0\\
 0              &\exp(2\pi ik/n)
\end{bmatrix}.
\end{equation}
Hence the representation $(\rho,W)$ of $\mathbb{Z}_n$ is given by: 
$(\rho,W)=\bigoplus_{k\in \mathbb{Z}_n}(\rho_k,E_k).$ This is the decomposition
of the representation of $\mathbb{Z}_n$ into irreducible representations.
Here, we denote $E=\bigsqcup_{k\in \mathbb{Z}_n}E_k \ (\text{where}\ \sqcup \ 
\text{denotes disjoint union of vector spaces})$ as the vector bundle over 
$\mathbb{Z}_n.$ 
The space $E_k$ is a section and we will denote the space of all sections by $W$ 
i.e. $W=\bigoplus_{k=0}^{k=n-1}E_{k}.$\\
\indent  There are altogether $n+1$ particles but the underlying symmetry (for 
all $n+1$ particles) is not $\mathbb{Z}_n$. More importantly $T\in \mathrm{Hom}(W,W)$ and 
hence we have to use the formula $\mathrm{tr}(TP_i)=\frac{1}{\#H}\sum_{k\in 
\mathbb{Z}_n}\chi_i(k)\mathrm{tr}\{t(0,k)\}$  with care. This is because in the
character formula $T$ is a homomorphism from $W$ to $W$ but the $T$ matrix we 
have in this problem is a homomorphism from $W\bigoplus E_{n}$ to
$W\bigoplus E_{n}$.\\
\indent We thus need to reformulation the nature of the symmetry group. This 
is done by introducing a new group 
$G=\{(n,j) \ \forall j\in \{0,1,2,\cdots,n-1\}\}$ which is isomorphic to $I\bigotimes \mathbb{Z}_n$. 
The elements of $G$  can be written as an ordered pairs $(n,j)$ where $n$ comes 
from the trivial group $I$ (containing only one element $n$) and the element $j$ comes
from the group $\mathbb{Z}_n$. Now we  attach with each element of $G$ a space isomorphic 
to $\mathbb{C}^2\bigoplus \mathbb{C}^2$ one for the $j^{th}$ particle and another for the $n^{th}$ particle. Under this new
formulation $E_{k}=\mathbb{C}^2\bigoplus \mathbb{C}^2\sim \mathbb{C}^4$ (i.e 
isomorphic to $\mathbb{C}^4$), $W=\bigoplus_{k=0}^{n-1}E_k$ and $T$ takes the 
following form;\\
$\mathcal{T}= $\\
$\scriptsize{
\tiny{\left[
\begin{array}{cc|cc|c|cc||cc|cc|c|cc}
i\omega & 0 & 0 & 0 & \cdots & 0 & 0
& 
I & 0 & 0 & 0 & \cdots & 0 & 0 \\
0 & 0 & 0 & 0 & \cdots & 0 & 0 
& 
0 & 1/mI & 0 & 0 & \cdots & 0 & 0\\
\hline
0 & 0 & i\omega & 0 & \cdots & 0 & 0
& 
0 & 0 & I & 0 & \cdots & 0 & 0 \\
0 & 0 & 0 & 0 & \cdots & 0 & 0
& 
0 & 0 & 0 & 1/mI & \cdots & 0 & 0\\
\hline
\vdots & \vdots & \vdots & \vdots & \ddots & \vdots & \vdots
& 
\vdots & \vdots & \vdots & \vdots & \ddots & \vdots & \vdots\\
\hline
0 & 0 & 0 & 0 & \cdots & i\omega & 0
& 
0 & 0 & 0 & 0 & \cdots & I & 0\\
0 & 0 & 0 & 0 & \cdots & 0 & 0
& 
0 & 0 & 0 & 0 & \cdots & 0 & 1/mI\\
\hline 
\hline
S_{00} & 0 & S_{01} & 0 & \cdots & S_{0n'} & S_{0n}
& 
i\omega & 0 & 0 & 0 & \cdots & 0 & 0\\
S_{0n} & 0 & S_{1n} & 0 & \cdots & S_{n'n} & S_{nn}
& 
0 & 0 & 0 & 0 & \cdots & 0 & 0\\
\hline
S_{10} & 0 & S_{11} & 0 & \cdots & S_{1n'} & S_{nn}
& 
0 & 0 & i\omega & 0 & \cdots & 0 & 0\\
S_{0n} & 0 & S_{1n} & 0 & \cdots & S_{n'n} & S_{nn}
& 
0 & 0 & 0 & 0 & \cdots & 0 & 0\\
\hline
\vdots & \vdots & \vdots & \vdots & \ddots & \vdots & \vdots
& 
\vdots & \vdots & \vdots & \vdots & \ddots & \vdots & \vdots\\
\hline
S_{n'0} & 0 & S_{n'1} & 0 & \cdots & S_{n'n'} & S_{n'n}
& 
0 & 0 & 0 & 0 & \cdots & i\omega & 0\\
S_{0n} & 0 & S_{1n} & 0 & \cdots & S_{n'n} & S_{nn}
& 
0 & 0 & 0 & 0 & \cdots & 0 & 0\\
\end{array}
\right]} }$
$=\begin{bmatrix}
  C  &  D\\
  P  &  Q 
 \end{bmatrix}
\text{where } n'=n-1$ and $I$ is the $2\times 2$ identity matrix.

From now on, by $T$ we mean the perturbation matrix $\begin{bmatrix}
                                            A&M^{-1}\\
                                            S&A
                                           \end{bmatrix}
$ and $\mathcal{T}$ as the newly formulated one.

In this new basis the first order linear perturbation equation takes the form; 
$$\begin{bmatrix}
   \left(\begin{array}{c}
\delta \dot{\textbf{x}}_0\\ 
   \delta \dot{\textbf{x}}_n
    \end{array}\right)\\
    \left(\begin{array}{c}
\delta \dot{\textbf{x}}_1\\ 
   \delta \dot{\textbf{x}}_n
    \end{array}\right)\\
    \vdots\\
    \left(\begin{array}{c}
\delta \dot{\textbf{x}}_{n-1}\\ 
   \delta \dot{\textbf{x}}_n
    \end{array}\right)\\
    \left(\begin{array}{c}
\delta \dot{\textbf{y}_0}\\ 
   \delta \dot{\textbf{y}}_n
    \end{array}\right)\\
    \left(\begin{array}{c}
\delta \dot{\textbf{y}}_1\\ 
   \delta \dot{\textbf{y}}_n
    \end{array}\right)\\
    \vdots\\
    \left(\begin{array}{c}
\delta \dot{\textbf{y}}_{n-1}\\ 
   \delta \dot{\textbf{y}}_n
    \end{array}\right)
  \end{bmatrix}
  =\mathcal{T}
  \begin{bmatrix}
   \left(\begin{array}{c}
\delta {\textbf{x}}_0\\ 
   \delta {\textbf{x}}_n
    \end{array}\right)\\
    \left(\begin{array}{c}
\delta {\textbf{x}}_1\\ 
   \delta {\textbf{x}}_n
    \end{array}\right)\\
    \vdots\\
    \left(\begin{array}{c}
\delta {\textbf{x}}_{n-1}\\ 
   \delta {\textbf{x}}_n
    \end{array}\right)\\
    \left(\begin{array}{c}
\delta {\textbf{y}_0}\\ 
   \delta {\textbf{y}}_n
    \end{array}\right)\\
    \left(\begin{array}{c}
\delta {\textbf{y}}_1\\ 
   \delta {\textbf{y}}_n
    \end{array}\right)\\
    \vdots\\
    \left(\begin{array}{c}
\delta {\textbf{y}}_{n-1}\\ 
   \delta {\textbf{y}}_n
    \end{array}\right)
  \end{bmatrix}
.$$ 
 Equation ~\eqref{EQUN1} tells us that a perturbation in the  
 position and momentum of a particular particle can be expressed as a linear 
combination of the  perturbed  positions and 
momenta of the other particles. \\
\indent Let us go back to equation ~\eqref{CF}.  We note that the left hand side of 
equation ~\eqref{CF} is the sum of different powers eigenvalues of the matrix $(TP_i)$. 
For our problem the value for $k=1,2,3,4$ give conditions for stability  that we need. 
This is because although we have 
taken into account the the presence of the center of mass in our formulation, it 
doesn't contribute anything to the character formula since 
 we have restricted ourselves to perturbations that do not change either the position
 or the momentum of the central mass. For this reason it is only
the four eigenvalues that describe the motion from equilibrium
of  particles on the ring. There is no motion of Saturn in the perturbation. 
The symmetry of the problem also tells us that since all the particles
 revolving about Saturn are identical it  follows that $\mathrm{tr}(TP_i)$ is the 
 same for all $i\in \{0,1,\cdots,n-1\}.$ \\
The linear stability of a system is analyzed by determining the 
eigenvalues of its perturbation matrix. In our problem stability requires that all the
 eigenvalues $\lambda_i$ of the perturbation matrix have to be purely imaginary. The 
 system, after perturbation, will then  oscillates about its original equilibrium 
configuration with a finite amplitude. This form of stability is appropriate for systems 
where dissipation of energy is not allowed. From the theory of equations 
we know that the purely imaginary roots of an algebraic 
equation of even order with real coefficients,must appear as complex conjugate pairs. 
It is easily checked that the perturbation matrix $\mathcal{T}$ of our problem
is of even order and there are at least four stability conditions, which are
\begin{enumerate} 
\item  $B^k=\sum\lambda_i^{k}=0$ (for $k=1,3$)
\item $B^2<0$
\item $B^4>0$.
\end{enumerate}
 These results hold when the four eigenvalues
of the perturbation matrix are all purely imaginary.
 It is easily checked that $B^k=0, k=1,3$. Hence a necessary condition 
for the stability is satisfied. Next we determine $\mathrm{tr}(\mathcal{T}P_0)$ and
check if the stability conditions holds for each irreducible subspace.

From equation ~\eqref{EQUN1} it is clear that 
$t(0,0)=t(0)=\begin{bmatrix}
              i\omega & 0 & I       & 0\\
              0       & 0 & 0       & 1/mI \\
              S_{00}  & 0 & i\omega & 0  \\
              S_{0n}  & 0 & 0       & 0
             \end{bmatrix}
,$\\
and \ \ \ $t(0,j)=t(j)=\begin{bmatrix}
                   0&0&0&0&\\
                   0&0&0&0&\\
                   S_{01}&0&0&0\\
                   S_{jn}&0&0&0
                  \end{bmatrix}  \text{for}  j\neq n \ \text{and} j\neq0
.$
This shows $\mathrm{tr}(TP_0)=0$ since $\mathrm{tr}(t(0))=\mathrm{tr}(t(j))=0$ i.e. $\sum\lambda_i=0$ over 
the space $E_0$ (remember in our notation $i=\begin{bmatrix}
                                                                                 
                                             0  & 1\\
                                                                                 
                                             -1 & 0
                                                                                 
                                            \end{bmatrix}
$). This is a much stronger stability condition. \\
The most general element $t(i,j)$ is given by $t(i,j)=\begin{bmatrix}
                                                        (C)_{ij} & (D)_{ij}\\
                                                        (P)_{ij} & (Q)_{ij}
                                                       \end{bmatrix}
.$
We have thus shown that the eigenvalues are of the matrix $\mathcal{T}P_{i}$ are 
purely imaginary and from this it follows, as stated before, that
$\sum\lambda_i^2<0.$  We next examine the consequence of this condition.\\
As discussed in the section on the character formula, $\mathrm{tr}(T^2P_{i})$ gives the 
sum of the squares of the eigenvalues. Squaring $T$ we get:
$T^2=\begin{bmatrix}
     A^2+M^{-1}S & AM^{-1}+M^{-1}A\\
     AS+SA       & SM^{-1}+A^2
    \end{bmatrix}.
$
Next we now calculate $\sum \lambda_i^2.$\\
First we reformulate $T^2$ in the new basis $$\scriptsize{{\begin{bmatrix}
   \left(\begin{array}{cccccccc}
\delta {\textbf{x}}_0\\ 
   \delta {\textbf{x}}_n
    \end{array}\right),&
    \hdots,&
    \left(\begin{array}{c}
\delta {\textbf{x}}_{n-1}\\ 
   \delta {\textbf{x}}_n
    \end{array}\right),&
    \left(\begin{array}{c}
\delta {\textbf{y}_0}\\ 
   \delta {\textbf{y}}_n
    \end{array}\right),&
    \hdots,&
    \left(\begin{array}{c}
\delta {\textbf{y}}_{n-1}\\ 
   \delta {\textbf{y}}_n
    \end{array}\right)
  \end{bmatrix}}^T}$$
  and then use the method discussed earlier to calculate the matrix $t(i,j)$
  and extract from it the matrix $t(0,j)$, where the identity  of the group
  has the label $0$.   Substituting  $t(0, j)$ in the character formula gives 
the relation 
  $\sum \lambda_i^2\propto \sum_{k\in \mathbb{Z}_n}\chi(k)\mathrm{tr}(t(0,k)),$
  we get;\\ 
  $\sum \lambda_i^2\propto 
[-4\omega^2+2m+4\times\frac{1}{4}\sum_{k=1}^{n-1}\csc^3(\frac{\theta_k}{2})]+\sum_{k=1
}^{n-1}4\cos(\theta_k)(-\frac{1}{4}\csc^3(\frac{\theta_k}{2})).$\\
 The sufficient condition for stability is  $\sum \lambda_i^2<0$ which gives; \\ 
$[-4\omega^2+2m+\sum_{k=1}^{n-1}\csc^3(\frac{\theta_k}{2})]+\sum_{k=1}^{n-1}
4\cos(\theta_k)(-\frac{1}{4}\csc^3(\frac{\theta_k}{2}))<0.$
 Substituting the value of $\omega^2$ in the above inequality we get;
 \begin{equation} \label{COND1} 
m>\frac{1}{2}\sum_{k=1}^{n-1}\csc^3(\frac{\theta_k}{2})-\frac{1}{2}\sum_{k=1}^{
n-1}\cos(\theta_k)\csc^3(\frac{\theta_k}{2})-2\sum_{k=1}^{n-1}\csc(\frac{
\theta_k}{2}).
\end{equation}
As the number of particles in the ring is very large we derive the stability 
condition for  the large $n$ limit.
Let us calculate the large $n$ limit of each term.~\cite{princeton}
 
$\sum_{k=1}^{n-1}\csc(\frac{\theta_k}{2})\longrightarrow\frac{2n}{\pi}\sum_{k=1
}^{(n-2)/2}\frac{1}{k}\approx\frac{2}{\pi}n\log(n/2),$\\
$\sum_{k=1}^{n-1}\csc^3(\frac{\theta_k}{2})\longrightarrow\frac{n^3}{\pi^3}
\sum_{k=1}^{[n-1]/2}\frac{1}{k^3}\approx\frac{2n^3}{\pi^3}\zeta(3) $\\
 where $\zeta$ denotes the Riemann zeta function and \\ 
 $\sum_{k=1}^{n-1}\csc^3(\theta_k/2)\cos\theta_k  \\ 
 \underset{large \ n}{\longrightarrow} \int_{2\pi/n}^{2\pi(n-1)/n}\csc^3(x/2)\cos(x)\frac{n}{2\pi}
dx$\\
 \ \ \ \ \ 
$=\frac{n}{2\pi}\left[\frac{1}{2}\csc^2(\pi/2n)+6\log\left(\frac{\tan(\pi/2n)}{
\cot(\pi/2n)}\right)-\sec^2(\pi/2n)\right]$\\
 \ \ \ \ \ $\underset{large \ 
n}{\longrightarrow}\frac{n}{2\pi}\left[\frac{1}{2}\left(\frac{4n^2}{\pi^2}
\right)-1\right].$\\
 Now substitution the above results in the relation ~\eqref{COND1} we get:
 
$m>\frac{n^3}{\pi^3}\left[\zeta(3)-\frac{1}{2}\right]+\underbrace{\left[\frac{n
}{4\pi}-\frac{1}{2}\log\left(\frac{n}{2}\right)\right]}_{\approx 0}$\\
$\Longrightarrow m>22.56\times 10^{-3}n^3.$\\
As we can see this value is smaller than the value found by Maxwell 
$(0.43n^3)$. But the reason for this is also clear because so far  we have only 
considered radial perturbation, we have left the angle between two
successive particles unchanged. This is not necessary and we can consider
more general perturbations. For example:
\begin{equation}\label{StPer}
\begin{bmatrix}
    \delta\textbf{x}\\
    \delta\textbf{y}
   \end{bmatrix}=\begin{bmatrix}
                 \widetilde{\rho} & 0\\
                 0 & \widetilde{\rho}
                \end{bmatrix}
                \begin{bmatrix}
                 \eta \\
                 \xi
                \end{bmatrix},
 \end{equation}               
  with $\widetilde{\rho}=\text{diag}(1,\rho,\rho^2,...,\rho^{n-1},1) \  
\text{and} \  \rho=e^{(2\pi i l/n)}$, where $l$ is a positive integer. Here
the angle differences are changed but the center of mass position is left
unchanged. Note that the angular position changes introduced still
use $Z_n$ symmetric variables.
The mass limit that we found earlier corresponds to the case $l=0$. The case 
$l=1$ corresponds to the situation when we group the identical particles 
in groups of two. Similarly 
$l=2$ corresponds to the configuration where the particles are grouped into 
groups of three. Accordingly we can have $l=3,4,\cdots$. We can clearly 
understand that the perturbation will be strongest when $l=(n-2)/2$. Any value 
greater than $(n-2)/2$ will repeat the situations for lower values of $l$. 
For the case $l=(n-2)/2$ we have actually grouped all the particles into two 
groups. The symmetry in this case is $\mathbb{Z}_2$. Now for a finite value 
of $n$ we can clearly understand that all integral values of $l$ are not possible.
For a given $n$,  the integer $l$ will 
take values equal to the factors of the number
$n$. But when we take the large $n$ limit, $l$ can practically take all possible 
values.\\
  \indent  Using this more general  perturbation condition we get; 
  \begin{equation}
   \begin{array}{rcl}
  \begin{bmatrix}
    \delta\dot{\textbf{x}}\\
    \delta\dot{\textbf{y}}
   \end{bmatrix} &=& 
   \begin{bmatrix}
   A&M^{-1}\\
   S&A
    \end{bmatrix}\begin{bmatrix}
                 \widetilde{\rho} & 0\\
                 0 & \widetilde{\rho}
                \end{bmatrix}
                \begin{bmatrix}
                 \eta \\
                 \xi
                \end{bmatrix}\\ 
                \\
           i.e. \      \begin{bmatrix}
                 \dot{\eta} \\
                 \dot{\xi}
                \end{bmatrix}&=& \begin{bmatrix}
                 \widetilde{\rho}^{-1} & 0\\
                 0 & \widetilde{\rho}^{-1}
                \end{bmatrix}\begin{bmatrix}
   A&M^{-1}\\
   S&A
    \end{bmatrix}\begin{bmatrix}
                 \widetilde{\rho} & 0\\
                 0 & \widetilde{\rho}
                \end{bmatrix}\begin{bmatrix}
                 \eta \\
                 \xi
                \end{bmatrix}
                =\widetilde{T}\begin{bmatrix}
                 \eta \\
                 \xi
                \end{bmatrix}
   \end{array}
  \end{equation}
where $\widetilde{T}=\begin{bmatrix}
                      \widetilde{\rho}^{-1}A\widetilde{\rho} & 
\widetilde{\rho}^{-1}M^{-1}\widetilde{\rho}\\
                      \widetilde{\rho}^{-1}S\widetilde{\rho} & 
\widetilde{\rho}^{-1}A\widetilde{\rho}
                     \end{bmatrix}
$.
 Thus we now have a different perturbation matrix 
  $\widetilde{\mathcal{T}}$ for our group $G$. Again we can see that 
  $\sum \lambda_i=0$ because $\mathrm{tr}(t(j))=0 \ \forall j\in\{0,1,\cdots,n-1\}$ hence 
 the necessary condition for stability is satisfied.
  Since the eigenvalues  are all purely imaginary we also have the 
conditions $\sum \lambda_i ^2<0$ and $\sum\lambda_i^4>0$ and so  on. We will comment on 
  the quartic constraint and possible higher order constraints later on.
  Now we show that the quadratic constraint leads to a mass constraint.
  
Substituting the relevant matrices in the character formula we get;\\
  $\sum \lambda_i ^2\varpropto 
\left[-4\omega^2+2m+4\frac{1}{4}\sum_{k=0}^{n-1}\csc^3(\frac{\theta_k}{2}
)\right]-\sum_{k=1}^{n-1}\cos(\theta_k)\cos(l\theta_k)\csc^3(\frac{\theta_k}{2}
)<0$ \\
i.e.
  \begin{equation}\label{COND2}
  \begin{array}{rcl}
m&>&\frac{1}{2}\sum_{k=1}^{n-1}\csc^3(\frac{\theta_k}{2}) \\
&-&\frac{1}{2}\sum_{k=1}^{n-1}\cos(\theta_k)\cos(l\theta_k)\csc^3(\frac{\theta_k}{2})\\
&-&2 \sum_{k=1}^{n-1}\csc(\frac{\theta_k}{2})
 \end{array}  
  \end{equation}
Thus we need to calculate 
$\sum_{k=1}^{n-1}\cos(\theta_k)\cos(l\theta_k)\csc^3(\frac{\theta_k}{2}).$\\
So we have;\\
$
\sum_{k=1}^{n-1}\cos(\theta_k)\cos(l\theta_k)\csc^3(\frac{\theta_k}{2})\\
=\frac{1}{2}\sum_{k=1}^{n-1}\left[\cos\{(l+1)\theta_k\}+\cos\{(l-1)\theta_k\}
\right]\csc^3(\theta_k/2)\\
=\sum_{k=1}^{(n-1)/2}\left[\cos\{(l+1)\theta_k\}+\cos\{(l-1)\theta_k\}\right]
csc^3(\theta_k/2)\\
=\sum_{k=1}^{(n-1)/2}\left[\cos\{(l+1)\frac{2\pi k}{n}\}+\cos\{(l-1)\frac{2\pi 
k}{n}\}\right]\csc^3\left(\frac{\pi k}{n}\right).
 $\\
 Substituting $l=(n-1)/2$ for the strongest perturbation and taking the large 
$n$ limit we get;\\
 $
 \begin{array}{rcl}
 \sum_{k=1}^{n-1}\cos(\theta_k)\cos(l\theta_k)\csc^3(\frac{\theta_k}{2})
 &\simeq& 2\sum\cos(\pi k)\csc^3\left(\frac{\pi k}{n}\right)\\
 &=&2\sum (-1)^k\csc^3\left(\frac{\pi k}{n}\right)\\
 &\simeq& 2\frac{n^3}{\pi^3}\sum_{k=0}^{\infty}\frac{(-1)^k}{k^3}\\
 &=&-\frac{3n^3}{2\pi^3}\sum_{k=0}^{\infty}\frac{1}{k^3}\\
 &=&-\frac{3n^3}{2\pi^3}\zeta(3).
 \end{array}
 $\\
 Substituting these results in ~\eqref{COND2} we get;
$$m>\frac{7n^3}{4\pi^3}\zeta (3)=0.068n^3.$$
 Thus the introduction of a perturbation that changes both
 the  radial positions and the angular coordinates in the system leads to
 a bigger mass value for stability.   However, it is 
 still smaller than the result found by Maxwell $(m>0.43n^3)$. But we have yet
 to consider other constraints on the eigenvalues identified that are present in the problem \\
 \indent Let us summarize the complete  set of eigenvalue constraints 
 necessary for  linear stability and hence justify the list
of constraints given before.  We note that 
$\mathcal{T}P_i$ is a $8\times 8$ matrix but the 
 effective rank of the matrix is 4 as the coupling between the two major 
$4\times4$
 matrices  of the system is very small.  Using
 this fact reduces the rank of the matrix $\mathcal{T}P_i$ to 4. So the 
characteristic polynomial of the matrix
 $\mathcal{T}P_i$ is going to be of order four in $\lambda$ and for such an 
equation there are four conditions for stability which have to
be imposed on the sums: $\sum\lambda_i,$ 
 $\sum\lambda_i^2,$ $\sum\lambda_i^3,$ and $\sum\lambda_i^4$. Out of these
 the sums of odd powers of the eigenvalues have to vanish.  These 
 two conditions we checked hold.  Then we have conditions on the sum
  of the square of the eigenvalues, which we have examined and finally  we 
  have the condition:  $\sum \lambda_i^4>0$ as a condition for stability. Besides 
these conditions  there is a further condition that follow from the  conditions listed.
  As the eigenvalues are purely imaginary and complex conjugate of each other
  we can write $\lambda_1=i\lambda$, $\lambda_2=-i\lambda$, $\lambda_3=i\mu$ and 
$\lambda_4=-i\mu$ where $\lambda$ and $\mu$ are positive constants.
 This gives;\\
  $\sum_{i=1}^{4}\lambda_i=0,$\\
  $\sum_{i=1}^{4}\lambda_i^2=-2(\lambda^2+\mu^2),$\\
  $\sum_{i=1}^{4}\lambda_i^3=0$ \\
  and $\sum_{i=1}^{4}\lambda_i^4=2(\lambda^4+\mu^4).$\\
  There no higher order independent constraints as the perturbation matrix is $4\times4$.
 Now we write $-2\lambda^2=x$, $-2\mu^2=y$ and $x+y=-\alpha$. Then, 
 $2(\lambda^2+\mu^2)=\frac{1}{2}(x^2+y^2)=\beta>0.$\\
 Thus we have
 $x+y=-\alpha \ \text{and} \ 
 x^2+y^2=2\beta.$
 Hence we get: $x^2+(-\alpha-x)^2=2\beta $
\begin{equation}
 \Longrightarrow x=-\frac{\alpha}{2}\pm\sqrt[]{2\beta}.
\end{equation}
 Now the condition that $x<0$ gives us a new condition to be satisfied:
 $2\beta<\alpha^2.$
We do not analyze these conditions further as our aim was to show
how easily the stability result $m>C n^3$ follows from group theory
and not to determine $C$.

\section{Conclusion}
We have described an induced representation  character formula method for
studying the stability of a system with a discrete group symmetry. The important 
point of the approach is that it does not require knowledge of the  invariant 
subspaces of the system. The entire procedure is group theoretical. We saw that
the reason for the emergence of induced representations was due to the fact that  
perturbations, in this framework, are linear maps between sections of a  
discrete vector bundle. Hence in order to exploit the symmetry properties of the 
bundle one needed to use induced representation.
The value of the constant $C$ in the stability condition 
$m>Cn^3$ we found by the group theoretical method was less than that found by Maxwell. 
This is  because we stopped at the level of the quadratic trace of eigenvalue constraints,
for instance,  the quartic constraint and other constraint identified, were 
not considered.  The  quartic constraint and other constraints when used lead 
to a quadratic equation in the square of one of the two imaginary eigenvalues of the system as shown in the other approaches and was thus not analyzed. Our  aim  was to show the power of the group theoretical approach to give the general structure of the mass constraint  in a very simple way. 
We believe the group theoretical method described is a useful 
and powerful tool for analyzing linear stability problems that have symmetry.
\subsection*{Acknowledgement}
Soumangsu Chakraborty would like to thank CRANN and Professor J.M.D.Coey
for a summer internship during which this work was carried out.

\bibliographystyle{unsrt}

\end{document}